# Constraints and relative entropies in nonextensive statistical mechanics

Sumiyoshi Abe [1,*] and G. B. Bagci [2]

[1] *Institute of Physics, University of Tsukuba, Ibaraki 305-8571, Japan*

[2] *Department of Physics, Middle East Technical University, Ankara 06531, Turkey*

**Abstract.** In nonextensive statistical mechanics, two kinds of definitions have been considered for expectation value of a physical quantity: one is the ordinary definition and the other is the normalized $q$-expectation value employing the escort distribution. Since both of them lead to the maximum-Tsallis-entropy distributions of a similar type, it is of crucial importance to determine which the correct physical one is. A point is that the definition of expectation value is indivisibly connected to the form of generalized relative entropy. Studying the properties of the relative entropies associated with these two definitions, it is shown how the use of the escort distribution is essential. In particular, the axiomatic framework proposed by Shore and Johnson is found to support the formalism with the normalized $q$-expectation value and to exclude the ordinary expectation value in nonextensive statistical mechanics.

## I. INTRODUCTION

Nonextensive statistical mechanics [1-4] pioneered by Tsallis [5] offers a consistent theoretical framework for the studies of complex systems in their nonequilibrium stationary states, systems with (multi)fractal and self-similar structures, long-range interacting systems, anomalous diffusion phenomena, and so on. The worked examples are dynamical systems at the edge of chaos [6-9], lattice Boltzmann models [10], magnetism of colossal magnetoresistance manganites [11], high-energy processes [12-16], cosmic rays [17], cellular aggregates [18], Lévy flights [19,20], semiclassical dynamic in optical lattices [21], astrophysics and self-gravitating systems [22,23], econophysical problems [24,25], kinetics of charged particles [26], Internet traffic [27], earthquakes [28,29], and complex networks [30-32]

In spite of these successes, still there remain some fundamental questions to be answered in the theory. One of them, which is of extreme importance, is concerned with the definition of expectation value. The most widely employed definition in nonextensive statistical mechanics is the normalized $q$-expectation value [33], which is given by

$$U^{(\text{nor})} = <H>_q = \sum_i P_i \, \varepsilon_i, \qquad (1)$$

$$P_i = \frac{(p_i)^q}{\sum_j (p_j)^q}, \qquad (2)$$

where $P_i$ is termed the escort distribution [34] associated with the basic distribution $p_i$ and $H$ denotes a physical random variable (e.g., the system energy) with its $i$th value $\varepsilon_i$. The index $q$ is taken to be positive. However, in the literature, there is an opinion that the ordinary expectation value

$$U^{(\text{ord})} = <H> = \sum_i p_i \varepsilon_i \qquad (3)$$

should be used also in nonextensive statistical mechanics. In Ref. [35], it has been shown that, for a class of power-law distributions, only the normalized $q$-expectation value is consistent with the method of steepest descents for (micro)canonical ensembles, but the situation remains unclear for the other class of distributions with compact supports.

Here, we address ourselves to the problem of choice of expectation value in nonextensive statistical mechanics. Our procedure is to examine the properties of the generalized relative entropies associated with the aforementioned two kinds of definitions. We shall see how the formalism with the normalized $q$-expectation value is superior to that with the ordinary expectation value.

The paper is organized as follows. In Sec. II, the ordinary and normalized $q$-expectation values are reexamined. A novel geometric aspect of the maximum entropy principle is also pointed out, there. In Sec. III, two different kinds of the generalized relative entropies associated with these definitions of expectation value are considered and their properties are studied. In Sec. IV, an axiomatic approach to the issue is developed, and is shown to support the normalized $q$-expectation value, excluding the possibility of using the ordinary expectation value in nonextensive statistical mechanics.

Sec. V is devoted to concluding remarks.

## II. ORDINARY AND NORMALIZED $q$-EXPECTATION VALUES

Before discussing the problem of expectation value, first we wish to point out a geometric aspect of the maximum entropy principle, which will be used later. The idea was inspired by the work of Bashkirov [36]. Consider a functional $\Phi$ defined in the space $\Sigma$ of probability distributions. Two operations on $\Sigma$ are of interest: one is translation and the other is dilatation. The corresponding generators are given by

$$T_i = \frac{\delta}{\delta p_i}, \tag{4}$$

$$D = \sum_i p_i \frac{\delta}{\delta p_i}, \tag{5}$$

respectively, where $p_i \in \Sigma$. Clearly, they satisfy the following closed algebra: $[T_i, T_j] = 0$, $[T_i, D] = T_i$, $[D, D] = 0$. Invariance of the functional $\Phi$ under these operations implies

$$T_i \Phi = 0, \tag{6}$$

$$D \Phi = 0. \tag{7}$$

If $\Phi$ is an entropic functional, the solution to these equations yields the maximum entropy distribution. A point to be noticed is that the dilatation operation is constrained by the normalization condition

$$\sum_i p_i - 1 = 0. \tag{8}$$

This, in turn, determines the value of the associated Lagrange multiplier.

Now, let us apply this method to the Tsallis entropy indexed by $q$ [5]

$$S_q[p] = \frac{1}{1-q}\left[\sum_i (p_i)^q - 1\right]. \tag{9}$$

Here and hereafter, the Boltzmann constant is set equal to unity for the sake of simplicity.

If the constraint is imposed on the ordinary expectation value, then the functional to be optimized is

$$\Phi^{(\text{ord})}[p; \alpha, \beta] = S_q[p] - \alpha\left(\sum_i p_i - 1\right) - \beta\left(\sum_i p_i \varepsilon_i - U^{(\text{ord})}\right), \tag{10}$$

where $\alpha$ and $\beta$ are the Lagrange multipliers. Eqs. (6)-(8) are found to give

$$\frac{q}{1-q}\left(\tilde{p}_i^{(\text{ord})}\right)^{q-1} - \alpha - \beta\varepsilon_i = 0, \tag{11}$$

$$\alpha = \frac{q}{1-q}\left[1 + (1-q)\tilde{S}_q^{(\text{ord})}\right] - \beta\tilde{U}^{(\text{ord})}, \tag{12}$$

where $\tilde{S}_q^{(\text{ord})}$ and $\tilde{U}^{(\text{ord})}$ are the values of $S_q$ and $U^{(\text{ord})}$ calculated in terms of the maximum entropy distribution $\tilde{p}_i^{(\text{ord})}$, respectively. From these equations, it follows that

$$\tilde{p}_i^{(\text{ord})} = \left[1 + (1-q)\tilde{S}_q^{(\text{ord})}\right]^{1/(q-1)}\left[1 - \frac{q-1}{q}\beta'\left(\varepsilon_i - \tilde{U}^{(\text{ord})}\right)\right]^{1/(q-1)}, \tag{13}$$

where

$$\beta' = \frac{\beta}{\sum_i \left(\tilde{p}_i^{(\text{ord})}\right)^q}. \tag{14}$$

On the other hand, if the normalized $q$-expectation value is employed, the corresponding functional reads

$$\Phi^{(\mathrm{nor})}[p;\alpha,\beta] = S_q[p] - \alpha\left(\sum_i p_i - 1\right) - \beta\left[\frac{\sum_i (p_i)^q \varepsilon_i}{\sum_j (p_j)^q} - U^{(\mathrm{nor})}\right]. \quad (15)$$

Here, we are using the same notation for the Lagrange multipliers as in Eq. (10), but it will not cause any confusion. The operator $D$ acting on the third term on the right-hand side trivially vanishes since this term is manifestly invariant under the dilatation. Eqs. (6)-(8) give rise to

$$\frac{q}{1-q}\left(\tilde{p}_i^{(\mathrm{nor})}\right)^{q-1} - \alpha - q\beta^*\left(\varepsilon_i - \tilde{U}^{(\mathrm{nor})}\right)\left(\tilde{p}_i^{(\mathrm{nor})}\right)^{q-1} = 0, \quad (16)$$

$$\alpha = \frac{q}{1-q}\left[1 + (1-q)\tilde{S}_q^{(\mathrm{nor})}\right], \quad (17)$$

where

$$\beta^* = \frac{\beta}{\sum_i \left(\tilde{p}_i^{(\mathrm{nor})}\right)^q}. \quad (18)$$

$\tilde{S}_q^{(\mathrm{nor})}$ and $\tilde{U}^{(\mathrm{nor})}$ are the values of $S_q$ and $U^{(\mathrm{nor})}$ calculated in terms of the maximum entropy distribution $\tilde{p}_i^{(\mathrm{nor})}$. Eqs. (16) and (17) lead to

$$\tilde{p}_i^{(\mathrm{nor})} = \frac{1}{\tilde{Z}_q^{(\mathrm{nor})}}\left[1 - (1-q)\beta^*\left(\varepsilon_i - \tilde{U}^{(\mathrm{nor})}\right)\right]^{1/(1-q)}, \quad (19)$$

$$\tilde{Z}_q^{(\mathrm{nor})} = \left[1 + (1-q)\tilde{S}_q^{(\mathrm{nor})}\right]^{1/(1-q)}$$

$$= \sum_i \left[1 - (1-q)\beta^*\left(\varepsilon_i - \tilde{U}^{(\mathrm{nor})}\right)\right]^{1/(1-q)}. \quad (20)$$

Eqs. (13) and (19) are quite similar to each other, but the signs of the exponents are opposite.

An important point is that with both of the definitions of expectation value the following thermodynamic relations hold:

$$\frac{\partial \tilde{S}_q^{(ord)}}{\partial \tilde{U}^{(ord)}} = \beta, \tag{21}$$

$$\frac{\partial \tilde{S}_q^{(nor)}}{\partial \tilde{U}^{(nor)}} = \beta, \tag{22}$$

which may indicate that the thermodynamic Legendre-transform structure is established in both cases. However, it is still an open problem in nonextensive statistical mechanics if $\beta$ is the physical inverse temperature [37].

It is clear that, in the limit $q \to 1$, the Tsallis entropy in Eq. (9) tends to the Boltzmann-Gibbs-Shannon entropy $S[p] = -\sum_i p_i \ln p_i$, and accordingly both of the distributions in Eqs. (13) and (19) converge to the familiar Boltzmann-Gibbs distribution $\tilde{p}_i \sim \exp(-\beta \varepsilon_i)$.

### III. GENERALIZED RELATIVE ENTROPIES

Relative entropy plays a fundamental role for comparing two distributions. There exist two different kinds of the generalized relative entropies in nonextensive statistical mechanics. One is given by [38]

$$I_q[p\|r] = \frac{1}{q-1} \sum_i p_i \left[ (p_i)^{q-1} - (r_i)^{q-1} \right] - \sum_i (p_i - r_i)(r_i)^{q-1}, \tag{23}$$

and the other is [39-41]

$$K_q[p\|r] = \frac{1}{1-q} \left[ 1 - \sum_i (p_i)^q (r_i)^{1-q} \right], \tag{24}$$

where $r_i$ is a reference distribution. In what follows, we shall look at the properties of these two quantities in detail. In particular, in subsection b, we shall see that $I_q[p\|r]$ and $K_q[p\|r]$ are the relative entropies associated with the ordinary expectation value and the normalized $q$-expectation value, respectively.

a.  **Correspondence relation and nonnegativity**

First of all, we notice that, in the limit $q \to 1$, both $I_q[p\|r]$ and $K_q[p\|r]$ tend to the Kullback-Leibler relative entropy

$$H[p\|r] = \sum_i p_i \ln \frac{p_i}{r_i}. \tag{25}$$

The following expression is often used for this quantity:

$$H[p\|r] = \frac{d}{dx} \sum_i (p_i)^x (r_i)^{1-x} \bigg|_{x \to 1}. \tag{26}$$

$K_q[p\|r]$ in Eq. (24) is obtained by replacing the differential operator by the Jackson $q$-differential operator [42]

$$K_q[p\|r] = D_q \sum_i (p_i)^x (r_i)^{1-x} \bigg|_{x \to 1}, \tag{27}$$

where $D_q f(x) = [f(qx) - f(x)] / [x(q-1)]$, which satisfies the $q$-deformed Leibniz rule $D_q[f(x)g(x)] = [D_q f(x)]g(x) + f(x) D_q[g(x)] + x(q-1) D_q[f(x)] D_q[g(x)]$ and converges to the ordinary differential in the limit $q \to 1$.

No such simple correspondence is known to exist between $H[p\|r]$ and $I_q[p\|r]$. One could, however, still dare to write

$$H[p\|r] = \frac{dG(x)}{dx} \bigg|_{x \to 1}, \tag{28}$$

$$G(x) = \frac{1}{x} \sum_i \left[ (p_i)^x - (r_i)^x \right] - \sum_i (p_i - r_i)(r_i)^{x-1}, \qquad (29)$$

and accordingly

$$I_q[p\|r] = q\, D_q\, G(x) \Big|_{x \to 1}. \qquad (30)$$

Like $H[p\|r]$, $I_q[p\|r]$ and $K_q[p\|r]$ are nonnegative and vanish if and only if $p_i = r_i$ ($\forall i$). This can be seen as follows. Regarding $I_q[p\|r]$, it is convenient to employ the integral representation [38]

$$I_q[p\|r] = \frac{q}{q-1} \sum_i \int_{r_i}^{p_i} ds \left[ s^{q-1} - (r_i)^{q-1} \right], \qquad (31)$$

from which nonnegativity follows immediately. On the other hand, $K_q[p\|r]$ can be rewritten as

$$K_q[p\|r] = \frac{1}{1-q} \sum_i p_i \left[ 1 - (r_i / p_i)^{1-q} \right]. \qquad (32)$$

Noticing $(1 - x^{1-q})/(1-q) \geq 1 - x$ for $x > 0$ and $q > 0$ with the equality for $x = 1$, $K_q[p\|r]$ is also seen to be nonnegative.

b.  **Free energy difference**

Let us discuss the physical meanings of $I_q[p\|r]$ and $K_q[p\|r]$, in particular, their relevance to the definitions of expectation value. For this purpose, we take the maximum entropy distributions as the reference distributions. The exponents of the maximum entropy distributions in Eqs. (13) and (19) together with the dependencies of $I_q[p\|r]$ and $K_q[p\|r]$ on $r_i$ should be noticed.

Putting $r_i = \tilde{p}_i^{(\text{ord})}$ in Eq. (23), we have

$$I_q[p\|\tilde{p}^{(\text{ord})}] = \beta\left(F_q^{(\text{ord})} - \tilde{F}_q^{(\text{ord})}\right), \tag{33}$$

where

$$F_q^{(\text{ord})} = U^{(\text{ord})} - \frac{1}{\beta}S_q, \qquad \tilde{F}_q^{(\text{ord})} = \tilde{U}^{(\text{ord})} - \frac{1}{\beta}\tilde{S}_q^{(\text{ord})}. \tag{34}$$

On the other hand, substituting $r_i = \tilde{p}_i^{(\text{nor})}$ into Eq. (24), we obtain

$$K_q[p\|\tilde{p}^{(\text{nor})}] = \frac{\hat{\beta}}{\sum_i \left(\tilde{p}_i^{(\text{nor})}\right)^q}\left(F_q^{(\text{nor})} - \tilde{F}_q^{(\text{nor})}\right), \tag{35}$$

where

$$\hat{\beta} = \beta^* \sum_i (p_i)^q, \tag{36}$$

$$F_q^{(\text{nor})} = U^{(\text{nor})} - \frac{1}{\hat{\beta}}S_q, \qquad \tilde{F}_q^{(\text{nor})} = \tilde{U}^{(\text{nor})} - \frac{1}{\hat{\beta}}\tilde{S}_q^{(\text{nor})}. \tag{37}$$

Eqs. (33) and (35) show that $I_q[p\|r]$ and $K_q[p\|r]$ are essentially the free energy differences and, therefore, are recognized as the generalized relative entropies associated with the ordinary expectation value and the normalized $q$-expectation value, respectively. We also mention that the quantum mechanical counterpart of $K_q[p\|r]$ has recently been employed to prove the second law of thermodynamics [43].

c. **Convexity**

Convexity is one of the most important properties to be fulfilled by relative entropy.

Taking the second-order derivatives of $I_q[p\|r]$ with respect to the arguments, one finds that it is convex in $p_i$, but *not* in $r_i$.

On the other hand, like the Kullback-Leibler relative entropy, $K_q[p\|r]$ is seen to be jointly convex (see Ref. [44] in the quantum mechanical case):

$$K_q\left[\sum_a \lambda_a p_{(a)} \middle\| \sum_a \lambda_a r_{(a)}\right] \leq \sum_a \lambda_a K_q[p_{(a)}\|r_{(a)}], \quad (38)$$

where $\lambda_a > 0$ and $\sum_a \lambda_a = 1$. This property is stronger than individual convexity in $p_i$ and $r_i$.

### d. Composability

Finally, we notice that, like the Kullback-Leibler relative entropy, $K_q[p\|r]$ is "composable" [45], but $I_q[p\|r]$ is not. In fact, for factorized joint distributions of a composite system $(A, B)$, $p_{ij}(A, B) = p_{(1)i}(A) \, p_{(2)j}(B)$ and $r_{ij}(A, B) = r_{(1)i}(A) \, r_{(2)j}(B)$, $K_q[p_{(1)} p_{(2)}\|r_{(1)} r_{(2)}]$ yields

$$K_q[p_{(1)}p_{(2)}\|r_{(1)}r_{(2)}] = K_q[p_{(1)}\|r_{(1)}] + K_q[p_{(2)}\|r_{(2)}]$$
$$+ (q-1) K_q[p_{(1)}\|r_{(1)}] \, K_q[p_{(2)}\|r_{(2)}], \quad (39)$$

whereas no such closed relation exists for $I_q[p_{(1)}p_{(2)}\|r_{(1)}r_{(2)}]$. Eq. (39) has its origin in the *q*-deformed Leibniz rule satisfied by the Jackson *q*-differential operator.

## IV. SHORE-JOHNSON AXIOMS

In the preceding section, we have seen how $K_q[p\|r]$ associated with the normalized *q*-expectation value has the properties, which are more favorable than those of $I_q[p\|r]$ corresponding to the ordinary expectation value. In this section, we discuss that the choice of $K_q[p||r]$ is in fact supported by a set of axioms.

About a quarter a century ago, Shore and Johnson [46] have proposed the axioms for minimum cross-entropy (i.e., relative entropy) principle. They are composed of the following five statements (presented in a nonabstract manner):

- *Axiom I* (Uniqueness): If the same problem is solved twice, then the same answer is expected to result both times.

- *Axiom II* (Invariance): The same answer is expected when the same problem is solved in two different coordinate systems, in which the posteriors in the two systems should be related by the coordinate transformation.

- *Axiom III* (System independence): It should not matter whether one accounts for independent information about independent systems separately in terms of their marginal distributions or in terms of the joint distribution.

- *Axiom IV* (Subset independence): It should not matter whether one treats independent subsets of the states of the systems in terms of their separate conditional distributions or in terms of the joint distribution.

- *Axiom V* (Expansibility): In the absence of new information, the prior (i.e., the reference distribution) should not be changed.

These axioms are extremely natural, and should be fulfilled in any physical situations.

For the Tsallis entropy in Eq. (9), the axioms and uniqueness theorem are known in the literature [47]. In contrast to this fact, the above set of axioms is quite general and not very restrictive, and therefore does not uniquely determine the definition of the relative entropy. However, still it can be shown [46] that the relative entropy $J[p\|r]$ with the prior $r_i$ and the posterior $p_i$ satisfying the axioms I-V has the following form:

$$J[p\|r] = \sum_i p_i\, h(p_i / r_i), \qquad (40)$$

where $h(x)$ is some function.

At this juncture, it is crucial to recognize that the function $h(x)$ surely exists for $K_q[p\|r]$:

$$h(x) = \frac{1}{1-q}(1 - x^{q-1}), \qquad (41)$$

whereas $I_q[p\|r]$ cannot be recast to the form in Eq. (40). Therefore, we conclude that the Shore-Johnson axioms support the normalized $q$-expectation value and exclude the possibility of using the ordinary expectation value in nonextensive statistical mechanics.

## V. CONCLUDING REMARKS

We have discussed two kinds of definitions of expectation value in nonextensive statistical mechanics, that is, the ordinary expectation value and the normalized $q$-expectation value. To determine which the correct definition is, we have studied the corresponding generalized relative entropies. It was found that the generalized relative entropy associated with the normalized $q$-expectation value has the properties, which are superior to those associated with the ordinary expectation value. More decisively, the Shore-Johnson axioms are shown to select the formalism with the normalized $q$-expectation value and to exclude the possibility of using the ordinary expectation value in nonextensive statistical mechanics.

## ACKNOWLEDGMENTS

S. A. thanks Andrei Bashkirov and Jan Naudts for discussions. He also thanks the Grant-in-Aid for Scientific Research of Japan Society for the Promotion of Science for financial support.